\documentclass[journal]{IEEEtran}
\usepackage{amsmath,epsfig,bm,amssymb,amsthm}
\usepackage{psfrag}
\usepackage{cite,color}
\usepackage{multirow}
\ifCLASSINFOpdf
\else
\fi

\usepackage{amsmath,epsfig,bm,amssymb,amsthm,balance}

\newcommand{\dd}{{\mathrm d}}

\newcommand{\bb}{{\mathrm b}}

\newcommand{\A}{\mathbf{A}}
\newcommand{\Lb}{\mathbf{L}}
\newcommand{\B}{\mathbf{B}}

\newcommand{\hb}{\mathbf{h}}
\newcommand{\Rb}{\mathbf{R}}
\newcommand{\Ib}{\mathbf{I}}
\newcommand{\y}{\mathbf{y}}
\newcommand{\w}{\mathbf{w}}
\newcommand{\diag}{\mbox{diag}}
\newcommand{\dgs}{\mbox{diag}\{\mathbf{s}\}}
\newcommand{\dgsc}{\mbox{diag}\{\mathbf{s}^*\}}
\newcommand{\s}{\mathbf{s}}
\newcommand{\U}{\mathbf{U}}

\newcommand{\Sgw}{\mathbf{\Sigma}_{\w}}
\newcommand{\E}{\mathrm{E}}
\newcommand{\C}{\mathbf{C}}
\newcommand{\CpN}{\mathbf{\mathcal{C}}^N}

\newcommand{\Q}{\mathbf{Q}}
\newcommand{\CN}{\mathcal{CN}}

\newcommand{\dgh}{\mbox{diag}\{\mathbf{h}_2\}}
\newcommand{\dghc}{\mbox{diag}\{\mathbf{h}_2^*\}}
\newcommand{\Sgy}{\mathbf{\Sigma}_{\y}}
\newcommand{\Sgb}{\mathbf{\Sigma}}

\newcommand{\Sg}{\mathbf{\Sigma}}

\newcommand{\opt}{\mathrm{opt}}

\newcommand{\bSigy}{\overline{\Sgb}_{\y}}
\newcommand{\MSD}{\mathrm{msd}}
\newcommand{\CDD}{\mathrm {cdd}}

\begin{document}
\title{Differential Dual-Hop Relaying under User Mobility
\thanks{The authors are with the Department of Electrical and Computer Engineering,
University of Saskatchewan, Saskatoon, Canada, S7N5A9.
Email: m.avendi@usask.ca, ha.nguyen@usask.ca.}}

\author{\IEEEauthorblockN{M. R. Avendi and Ha H. Nguyen}}
%Department of Electrical and Computer Engineering\\
%University of Saskatchewan, Saskatoon, Canada, S7N5A9\\
%Email: m.avendi@usask.ca, ha.nguyen@usask.ca.
%}

\maketitle

\begin{abstract}
\label{abs}
This paper studies dual-hop amplify-and-forward relaying system employing differential encoding and decoding over time-varying Rayleigh fading channels. First, the convectional ``two-symbol'' differential detection (CDD) is theoretically analysed in terms of the bit-error-rate (BER). {The obtained analysis clearly shows that performance of two-symbol differential detection severely degrades in fast-fading channels and reaches an irreducible error floor at high signal-to-noise ratio region. To overcome the error floor experienced with fast-fading, a practical suboptimal ``multiple-symbol'' detection (MSD) is designed and its performance is theoretically analysed.} The analysis of CDD and MSD are verified and illustrated with simulation results under different fading scenarios. {Specifically, the obtained results show that the proposed MSD can significantly improve the system performance in fast-fading channels.}
\end{abstract}

\begin{keywords}
Dual-hop relaying, amplify-and-forward, differential $M$-PSK, non-coherent detection, time-varying fading channels, multiple-symbol detection.
\end{keywords}

\IEEEpeerreviewmaketitle

\section{Introduction}
\label{se:intro}
Dual-hop relaying without a direct link has been considered in the literature as a technique to leverage coverage problems of many wireless applications such as 3GPP LTE, WiMAX, WLAN, Vehicle-to-Vehicle communication and wireless sensor networks \cite{coop-deploy,coop-V2V,coop-WiMAX,coop-dohler,coop-LTE}. Such a technique can be seen as a type of cooperative communication in which one node in the network helps another node to communicate with (for example) the base station when the direct link is very poor or the user is out of the coverage area.

A two-phases transmission process is usually utilized in such a network. Here, Source transmits data to Relay in the first phase, while in the second phase Relay performs amplify-and-forward (AF) strategy to send the received data to Destination \cite{coop-laneman}. {For future reference in this paper, the overall channel from Source, through Relay, to Destination is refereed to as the cascaded channel.} Error performance of dual-hop relaying without direct link employing AF strategy has been studied in \cite{dual-hop-Hasna,dh-smith09,dh-hasna03}. Also, the statistical properties of the cascaded channel between Source and Destination in a dual-hop AF relaying have been examined in \cite{SPAF-P}.

In the existing literature, either coherent detection or \emph{slow-fading} environment is assumed. In coherent detection, instantaneous channel state information (CSI) of both Source-Relay and Relay-Destination links are required at Destination. This requirement, and specifically Source-Relay channel estimation, would be challenging to meet for some applications. Also, when Source and/or Relay are mobile, the constructed channels become time-varying. On one hand, the channel variation makes coherent detection inefficient or sometimes impossible due to the requirement of fast channel estimation or tracking. On the other hand, by employing differential modulation and two-symbol non-coherent detection, this variation can be somewhat tolerated as long as the channel does not change significantly over two consecutive symbols. { For the case of conventional differential detection (CDD) using two symbols, performance of AF relay networks in slow-fading channels has been considered in \cite{DAF-General,DAF-GLAT,DAF-MN-Himsoon,DAF-Liu,DAF-DDF-QZ}. It has been shown that over slow-fading channels, there exists about 3 dB performance gap between coherent and non-coherent detection schemes.} However, in practical time-varying channels, the effect of channel variation can lead to a much larger degradation.

Motivated from the above discussion, the first goal in this paper is to analyse the performance of single-branch dual-hop relaying employing differential $M$-PSK and two-symbol non-coherent detection in time-varying Rayleigh fading channels. We refer to this system as differential dual-hop (D-DH) relaying. In \cite{DAF-ITVT}, we studied a multi-branch differential AF relaying with direct link in time-varying channels, however only a lower bound of the bit-error rate (BER) was derived for the multi-branch system. Although, the dual-hop relaying considered in this paper is a special case of multi-branch relaying, the analysis here is different than that of \cite{DAF-ITVT}. Specifically, for the case of two-symbol non-coherent detection, an exact bit error rate (BER) expression is obtained. It is also shown that the system performance quickly degrades in fast-fading channels and reaches an error floor at high transmit power. The theoretical value of the error floor is also derived and used to investigate the fading rate threshold. Interestingly, it is seen that the error floor only depends on the auto-correlation value of the cascaded channel and modulation parameters.

%Counterpart to point-to-point communications \cite{msdd-div,msdd-div2,MSDSD-L,msdd_fung,MSDUSTC-P,msd-Rick},
The second goal in this paper is to design a multiple-symbol detection (MSD) for the D-DH relaying to improve its performance in fast-fading channels. Multiple-symbol detection has been considered for point-to-point communications in \cite{msdd-div,msdd-div2,MSDSD-L,msdd_fung,MSDUSTC-P,msd-Rick}. The challenge in developing multiple-symbol detection for AF relay networks is that, due to the complexity of the distribution of the received signal at Destination, the optimum decision metric does not yield a closed form solution. To circumvent this problem, here, the optimum decision rule is replaced with an alternative decision rule and further simplified to be solved with low complexity. Furthermore, theoretical error performance of the proposed MSD is obtained. This analysis is useful to investigate a trade-off between the MSD window size and the desired performance. The error analysis of both  CDD and MSD is thoroughly verified with simulation results in various fading scenarios.

{The obtained results will show that while CDD performs quite well in slow-fading channels, it fails to provide satisfactory performance in fast-fading channels. On the other hand, it will be seen that the proposed MSD detection is very robust against channel variation and can yield significant performance improvement over the CDD. Obviously, such an improvement comes at the price of higher detection complexity. Nevertheless, using sphere decoding techniques applicable to the proposed MSD detection could reduce the computation complexity. Since coherent detection is not useful at all in fast-fading channels, the proposed multiple-symbol differential detection would be the only alternative.}

{The main contributions of this paper are summarized as follows. The exact theoretical BER analysis of two-symbol differential detection for a dual-hop relay network in time-varying Rayleigh-fading channels is derived and verified with simulation results in various fading scenarios. The existence of an error floor when using two-symbol differential detection in high signal-to-noise ratio (SNR) region can be predicted and its value is theoretically derived and verified by simulation results. A low-complexity suboptimal  multiple-symbol differential detection is developed for dual-hop relaying to improve the poor performance of two-symbol differential detection in fast-fading channels. Performance of the proposed multiple-symbol differential detection is also theoretically analyzed and verified by simulation results in various fading scenarios.}

The outline of the paper is as follows. Section \ref{sec:system} describes the system model. In Section \ref{sec:two-symbol}, two-symbol differential detection and its performance over time-varying channels are studied. Section \ref{sec:MSDSD} develops the MSD algorithm and analyses its performance. Simulation results are given in Section \ref{sec:sim}. Section \ref{sec:con} concludes the paper.

\emph{Notation}: Bold upper-case and lower-case letters denote matrices and vectors, respectively. $(\cdot)^t$, $(\cdot)^*$, $(\cdot)^H$ denote transpose, complex conjugate and Hermitian transpose of a complex vector or matrix, respectively. $|\cdot|$  denotes the absolute value of a complex number and $\|\cdot \|$ denotes the Euclidean norm of a vector. $\mathcal{CN}(0,N_0)$ stands for complex Gaussian distribution with zero mean and variance $N_0$. $\mbox{E}\{\cdot\}$ denotes expectation operation. Both ${e}^{(\cdot)}$ and $\exp(\cdot)$ show the exponential function. $\dgs$ is the diagonal matrix with components of $\s$ on the main diagonal and $\Ib_N$ is the $N \times N$ identity matrix. A symmetric $N\times N$ Toeplitz matrix is defined by $\mbox{toeplitz}\{x_1,\cdots,x_N\}$. $\mbox{det}\{\cdot\}$ denotes determinant of a matrix. $\CpN$ is the set of complex vectors with length $N$. $\Re\{\cdot\}$ and $\Im\{\cdot\}$ denote the real and imaginary parts of a complex number.

\section{System Model}
\label{sec:system}
\begin{figure}[t]
\psfrag {Source} [] [] [1.0] {Source}
\psfrag {Relay} [] [] [1.0] {Relay}
\psfrag {Destination} [] [] [1.0] {Destination}
\psfrag {h1} [] [] [1.0] {$\qquad h_1[k]$\;\;\;}
\psfrag {h2} [] [] [1.0] {$\qquad h_2[k]$}
\centerline{\epsfig{figure={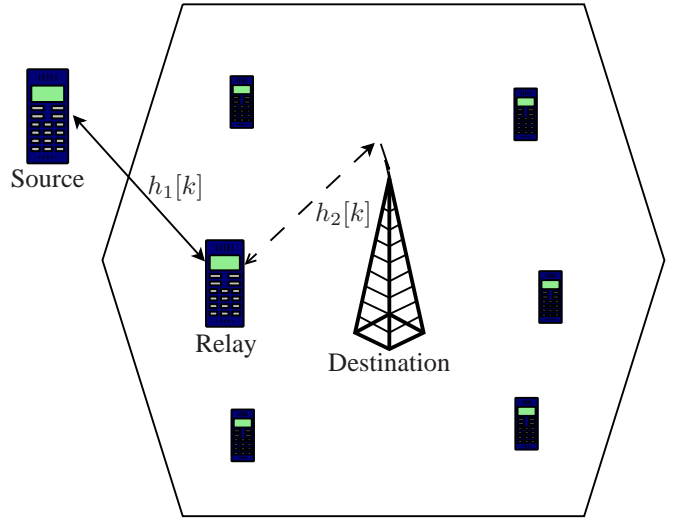},width=8.5cm}}
\caption{Illustration of single-branch dual-hop relaying system without direct link.}
\label{fig:sysmodel}
\end{figure}
The wireless relay model under consideration, depicted in Figure~\ref{fig:sysmodel}, has one Source, one Relay and one Destination. Source is out of the cell coverage and hence the received signal in the direct link is not sufficiently strong to facilitate data transmission. Therefore, with the help of another user (Relay), a dual-hop relaying system without direct link is constructed to connect Source to Destination. Each node has a single antenna, and the communication between nodes is half duplex (i.e., each node is able to only send or receive in any given time). The channels from Source to Relay (SR) and from Relay to Destination (RD) are denoted by $h_1[k]$ and $h_2[k]$, respectively, where $k$ is the symbol time. A Rayleigh flat-fading model is assumed for each channel, i.e., $h_1[k]\sim \CN(0,\sigma_1^2)$ and $h_2[k]\sim \CN(0,\sigma_2^2)$. The channels are spatially uncorrelated and changing continuously in time. The time-correlation between two channel coefficients, $n$ symbols apart, follows the Jakes' model \cite{microwave-jake}:
\begin{equation}
\varphi_i(n)=\E \{h_i[k]h_i^*[k+n]\}=\sigma_i^2 J_0(2\pi f_i n),\quad i=1,2
\end{equation}
where $J_0(\cdot)$ is the zeroth-order Bessel function of the first kind and $f_i$ is the maximum normalized Doppler frequency of the $i$th channel. {For ease of discussion, the overall channel from Source, through Relay, to Destination shall be called the cascaded channel.}

At time $k$, a group of $\log_2M$ information bits is mapped to a $M$-PSK symbol as $v[k]\in \mathcal{V}$ where $\mathcal{V}=\{e^{j2\pi m/M},\; m=0,\dots, M-1\}$. Before transmission, the symbols are encoded differentially as
\begin{equation}
\label{eq:s-source}
s[k]=v[k] s[k-1],\quad s[0]=1.
\end{equation}
The transmission process is divided into two phases. Block-by-block transmission protocol is utilized to transmit a frame of symbols in each phase as symbol-by-symbol transmission causes frequent switching between reception and transmission, which is not practical. However, the analysis is the same for both cases and only the channel auto-correlation value is different ($n=1$ for block-by-block and $n=2$ for symbol-by-symbol).

{Information can be transmitted from Source to Destination (uplink) or from Destination to Source (downlink). Without loss of generality, here the uplink transmission is assumed.} In phase I, the symbol $\sqrt{P_0}s[k]$ is transmitted by Source to Relay, where $P_0$ is the average source power per symbol. The received signal at Relay is
\begin{equation}
\label{eq:relay_rx}
x[k]=\sqrt{P_0}h_1[k]s[k]+w_1[k]
\end{equation}
where $w_1[k]\sim \CN(0,N_0)$ is the noise component at Relay. Also, the average received SNR per symbol at Relay is defined as
\begin{equation}
\label{eq:rho1}
\rho_1=\frac{P_0\sigma_1^2}{N_0}.
\end{equation}

The received signal at Relay is then multiplied by an amplification factor $A$, and re-transmitted to Destination. The amplification factor, based on the variance of SR channel, is commonly used in the literature as
\begin{equation}
\label{eq:A}
A =\sqrt{\frac{P_1}{P_0\sigma_1^2+N_0}},
\end{equation}
to normalize the average power per symbol at Relay to $P_1$. However,
$A$ can be any arbitrarily fixed value. The corresponding received signal at Destination is
\begin{equation}
\label{eq:dest-rx1}
y[k]=Ah_2[k]x[k]+w_2[k],
\end{equation}
where $w_2[k]\sim \CN(0,N_0)$ is the noise component at Destination. Substituting (\ref{eq:relay_rx}) into (\ref{eq:dest-rx1}) yields
\begin{equation}
\label{eq:Destination-rx}
y[k]= A \sqrt{P_0}h[k]s[k]+w[k],
\end{equation}
where $h[k]=h_1[k]h_2[k]$ is the cascaded channel with zero mean and variance $\sigma_1^2\sigma_2^2$ \cite{SPAF-P}, and
\begin{equation}
\label{eq:w[k]}
w[k]=A h_2[k]w_1[k]+w_2[k]
\end{equation}
is the equivalent noise at Destination.
It should be noted that for a given $h_2[k]$, $w[k]$ is a complex Gaussian random variable with zero mean and variance
\begin{equation}
\label{eq:sig_wk}
\sigma_{w}^2=N_0 \left(1+ A^2 |h_2[k]|^2\right).
\end{equation}
Thus $y[k]$, conditioned on $s[k]$ and $h_2[k]$, is a complex Gaussian random variable as well.

In the following section, the conventional two-symbol differential detection (CDD) of the received signals at Destination and its performance are considered.

\section{Two-Symbol Differential Detection}
\label{sec:two-symbol}

\subsection{Detection Process}
\label{subsec:channel-model}
Given two consecutive received symbols at a time, non-coherent detection of the transmitted symbol is obtained by the following minimization:
\begin{equation}
\label{eq:ml-detection}
\hat{v}[k]= \arg \min \limits_{v[k]\in \mathcal{V}} |y[k]- v[k] y[k-1]|^2.
\end{equation}
As can be seen, no channel information is needed for detection. Two-symbol detection is simple to implement and it is mainly based on the assumption that the channel coefficients are approximately constant during two adjacent symbols. Although this assumption would be true for slow-fading channels, it would be violated when users are fast moving. In the next section, the performance of two-symbol non-coherent detection in time-varying Rayleigh fading channel is analysed.

\subsection{Performance Analysis}
\label{sec:symbol_error_probability}
Using the unified approach in \cite[eq.25]{unified-app}, it follows that the conditional BER for two-symbol differential detection can be written as
\begin{equation}
\label{eq:Pb-gama-hrd}
P_{\bb}^{\CDD}(E|\gamma,h_2)=\frac{1}{4\pi} \int \limits_{-\pi}^{\pi} g(\theta) e^{-q(\theta)\gamma} \dd \theta
\end{equation}
where $g(\theta)=(1-\beta^2)/(1+2\beta\sin(\theta)+\beta^2)$, $q(\theta)=(b^2/\log_2 M) (1+2\beta\sin(\theta)+\beta^2)$, and $\beta=a/b$. The values of $a$ and $b$ depend on the modulation size \cite{unified-app}. Also, $\gamma$ is the instantaneous effective SNR at the output of the differential detector which needs to be determined for time-varying channels.

To proceed with the performance analysis of two-symbol differential detection in time-varying channels, it is required to model the time-varying nature of the channels. For this purpose, individual Rayleigh-faded channels, i.e., Source-Relay and Relay-Destination channels, are expressed by a first-order auto-regressive (AR(1)) model as
\begin{gather}
\label{eq:ARi}
h_i[k]=\alpha_i h_i[k-1]+\sqrt{1-\alpha_i^2} e_i[k],\quad i=1, 2
\end{gather}
where $\alpha_i=\varphi_i(1)/\sigma_i^2$ is the auto-correlation of the $i$th channel and $e_i[k]\sim \mathcal{CN}(0,\sigma_i^2)$ is independent of $h_i[k-1]$. Based on these expressions, a first-order time-series model has been derived in \cite{DAF-ITVT} to characterise the evolution of the cascaded channel in time. The time-series model of the cascaded channel is given as (the reader is referred to \cite{DAF-ITVT} for the detailed derivations/verification)
\begin{equation}
\label{eq:ARmodel}
h[k]=\alpha h[k-1]+\sqrt{1-\alpha^2}\ h_2[k-1]e_1[k]
\end{equation}
where $\alpha=\alpha_1 \alpha_2 \leq 1$ is the equivalent auto-correlation of the cascaded channel, which is equal to the product of the auto-correlations of individual channels, and $e_1[k]\sim \mathcal{CN}(0,\sigma_1^2)$ is independent of $h[k-1]$.

By substituting (\ref{eq:ARmodel}) into (\ref{eq:Destination-rx}) one has
\begin{equation}
\label{eq:ykk-1}
y[k]=\alpha v[k]y[k-1]+n[k],
\end{equation}
where
\begin{multline}
\label{eq:n[k]}
n[k]=w[k]- \alpha v[k]w[k-1]
\\+ \sqrt{1-\alpha^2}A\sqrt{P_0} s[k] h_2[k-1] e_1[k].
\end{multline}
From expression \eqref{eq:ykk-1}, with given $y[k]$ and $y[k-1]$, the non-coherent detection process can be interpreted as coherent detection of data symbol $v[k]$ distorted by a fading channel equivalent to $y[k-1]$ and in the presence of the equivalent noise $n[k]$.
%As can be seen from the model in \eqref{eq:n[k]}, the exact distribution of $n[k]$ is difficult to find. However, by substituting $h_2[k]$ from \eqref{eq:ARi} into $w[k]$ one has
%\begin{multline}
%\label{eq:w[k]-hat}
%w[k]=w_2[k]+A \alpha_2 w_1[k]  h_2[k-1]\\+A\sqrt{1-\alpha_2^2}w_1[k] e_2[k]
%\approx w_2[k]+A \alpha_2 w_1[k]  h_2[k-1]
%\end{multline}
%where the approximation comes from the observation that even for very fast-fading channels the term $A\sqrt{1-\alpha_2^2}$ is very small. Hence, using the approximated value of $w[k]$ into \eqref{eq:n[k]},
%\begin{multline}
%\label{eq:n[k]-app}
%n[k]\approx w_2[k]+A\alpha_2 w_1[k] h_2[k-1]\\
%-\alpha v[k] \left( w_2[k-1]+Ah_2[k-1]w_1[k-1] \right)\\
%+\sqrt{1-\alpha^2} A \sqrt{P_0} s[k] h_2[k-1] e_1[k]
%\end{multline}
%which shows that, conditioned on $h_2[k-1]$, $n[k]$ is a linear combination of complex Gaussian random variables and hence it is also complex Gaussian random variable.
%From now on, the time index $[k-1]$ is omitted in this section to simplify the notation.
Hence, for time-varying channels, based on \eqref{eq:ykk-1} and \eqref{eq:n[k]}, $\gamma$ is computed as
\begin{equation}
\label{eq:gama_d}
\gamma=\bar{\gamma} |h_1|^2
\end{equation}
where
\begin{equation}
\label{eq:gama_b}
\overline{\gamma}=\frac{\alpha^2 A^2 (P_0/N_0) |h_2|^2}{1+\alpha^2+\left[1+\alpha^2+(1-\alpha^2)\rho_1 \right] A^2|h_2|^2}.
\end{equation}
In the above, the time index $[k-1]$ is omitted to simplify the notation. Clearly, for slow-fading channels ($\alpha=1$), the equivalent noise power is only enhanced by a factor of two and $\gamma$ is half of the received SNR in coherent detection $A^2 (P_0/N_0) |h_2|^2 |h_1|^2/(1+A^2|h_2|^2)$ \cite{dual-hop-Hasna,SPAF-P}, as expected. However, for fast-fading channels, $\alpha<1$, the noise power is dominated by the last term in \eqref{eq:n[k]} and then significantly increases with increasing transmit power. This leads to a larger degradation in the effective SNR and poor performance of two-symbol non-coherent detection in fast-fading channels.

Since, $\lambda_1=|h_1|^2$ is exponentially distributed, i.e., $f_{\lambda_1}(\lambda)=\exp(-\lambda/\sigma_1^2)/\sigma_1^2$, the variable $\gamma$, conditioned on $|h_2|$, follows an exponential distribution with the following pdf and cdf:
\begin{equation}
\label{eq:pdf-gama_b}
f_{\gamma|h_2}(\gamma)=\frac{1}{\overline{\gamma}\sigma_1^2} \exp\left(-\frac{\gamma}{\overline{\gamma}\sigma_1^2}\right)
\end{equation}
\begin{equation}
\label{eq:cdf-gama_b}
F_{\gamma|h_2}(\gamma)=1- \exp\left(-\frac{\gamma}{\overline{\gamma}\sigma_1^2}\right).
\end{equation}

By substituting $\gamma$ into \eqref{eq:Pb-gama-hrd} and taking the average over the distribution of $\gamma$, one has
\begin{equation}
\label{eq:Pb-hrd}
P_{\bb}^{\CDD}(E|h_2)=\frac{1}{4\pi} \int \limits_{-\pi}^{\pi} g(\theta) I(\theta) \dd \theta
\end{equation}
where
\begin{multline}
\label{eq:I_theta}
I(\theta)=\int \limits_{0}^{\infty} e^{-q(\theta)\gamma} \frac{1}{\overline{\gamma}\sigma_1^2} e^{-\frac{\gamma}{\overline{\gamma}\sigma_1^2}} \dd \gamma=\frac{1}{\overline{\gamma}\sigma_1^2 q(\theta)+1}\\
=b_3(\theta) \frac{\lambda_2+b_1}{\lambda_2+b_2(\theta)}
\end{multline}
with $\lambda_2=|h_2|^2$, $b_3(\theta)=b_2(\theta)/b_1$ and $b_1$, $b_2(\theta)$ defined as
\begin{align*}
\label{eq:b1b2b3}
b_1=\frac{1+\alpha^2}{(1+\alpha^2)A^2+(1-\alpha^2)A^2\rho_1}\\
b_2(\theta)=\frac{1+\alpha^2}{(1+\alpha^2)/b_1+\alpha^2 q(\theta)A^2\rho_1}
\end{align*}
%The later expression in \eqref{eq:I_theta} is obtained by substituting $\bar{\gamma}$.

Now, by taking the final average over the distribution of $\lambda_2=|h_2|^2$, $f_{\lambda_2}(\lambda)=\exp\left(-\lambda/\sigma_2^2\right)/\sigma_2^2$, it follows that
\begin{equation}
\label{eq:Pb}
P_{\bb}^{\CDD}(E)=\frac{1}{4\pi} \int \limits_{-\pi}^{\pi} g(\theta) J(\theta) \dd\theta
\end{equation}
where
\begin{multline}
\label{eq:J_theta}
J(\theta)=\int \limits_{0}^{\infty} b_3(\theta) \frac{\lambda+b_1}{\lambda+b_2(\theta)} \frac{1}{\sigma_2^2} e^{\left(-\frac{\lambda}{\sigma_2^2}\right)} \dd\lambda\\
=b_3(\theta) \left( 1+\frac{1}{\sigma_2^2}(b_1-b_2(\theta)) e^{\left(\frac{b_2(\theta)}{\sigma_2^2}\right)} E_1\left(\frac{b_2(\theta)}{\sigma_2^2}\right) \right)
\end{multline}
and $E_1(x)=\int \limits_x^{\infty} (e^{-t}/t) \dd t$ is the exponential integral function. The definite integral in \eqref{eq:Pb} can be easily computed using numerical methods and it gives an exact value of the BER of the D-DH system under consideration in time-varying Rayleigh fading channels.

It is also informative to examine the expression of $P_{\bb}(E)$ at high transmit power. In this case,
\begin{equation}
\label{eq:gb_limit}
\lim \limits_{(P_0/N_0)\rightarrow \infty} E[\bar{\gamma}]= \frac{\alpha^2}{(1-\alpha^2)\sigma_1^2},
\end{equation}
which is independent of $|h_2|^2$ and $(P_0/N_0)$. Therefore, by substituting the converged value into (\ref{eq:Pb-hrd}), the error floor appears as
\begin{equation}
\label{eq:PEP-floor}
\lim \limits_{(P_0/N_0) \rightarrow \infty} P_{\bb}^{\CDD}(E)= \frac{1}{4\pi} \int \limits_{-\pi}^{\pi} g(\theta) \frac{1-\alpha^2}{\alpha^2q(\theta)+1-\alpha^2} \dd \theta.
\end{equation}
It is seen that the error floor can be determined based on the amount of the equivalent channel auto-correlation (i.e., parameter $\alpha$) and also the parameters of the signal constellation (i.e., $g(\theta)$ and $q(\theta)$). Thus, one way to control this error floor would be to keep the normalized Doppler frequency as low as possible by reducing the symbol duration of the system.

\section{Multiple-Symbol Detection}
\label{sec:MSDSD}
As discussed in the previous section, two-symbol non-coherent detection suffers from a high error floor in fast-fading channels. To overcome such a limitation, this section designs and analyses a multiple-symbol detection scheme that takes a window of the received symbols at Destination for detecting the transmitted signals. %The detection process is still non-coherent, i.e., no instantaneous CSI is needed, though, the second order statistics of the channels (variance and auto-correlation) are required for the detection process.

\subsection{Detection Process}
Let the $N>2$ received symbols be collected in vector $\y=\left[\; y[1],y[2],\dots, y[N]\; \right]^t$, which can be written as
\begin{equation}
\label{eq:Y}
\y=A\sqrt{P_0} \dgs \diag\{\hb_2\} \hb_1 +\w
\end{equation}
where
\begin{gather}
\s= \left[\; s[1],\cdots, s[N]\; \right]^t \\
\hb_2=\left[\; h_2[1],\cdots, h_2[N] \;\right]^t \\
\hb_1=\left[\; h_1[1],\cdots, h_1[N]\; \right]^t \\
\w=\left[\; w[1],\cdots, w[N]\; \right]^t.
\end{gather}
Therefore, conditioned on both $\s$ and $\hb_2$, $\y$ is a circularly symmetric complex Gaussian vector with the following pdf:
\begin{equation}
\label{eq:pdfY}
P(\y|\s,\hb_2)=\frac{1}{\pi^N \mathrm{det}\{\Sgy\}} \exp\left( -\y^H \Sgy^{-1} \y \right).
\end{equation}
In \eqref{eq:pdfY}, the matrix $\Sgy$ is the conditional covariance matrix of $\y$, defined as
\begin{equation}
\label{eq:RY}
\begin{split}
\Sgy=&\E \{ \y \y^H | \s,\hb_2 \}=\\
&A^2 P_0 \dgs \diag\{\hb_2\} \Sg_{\hb_1} \diag\{\hb_2^*\} \dgsc+\Sgw
\end{split}
\end{equation}
with
\begin{gather}
\Sg_{\hb_1}=\E\{ \hb_1 \hb_1^H \}=\mathrm{toeplitz}\left\lbrace \varphi_1(0),\dots,\varphi_1(N-1)\right\rbrace, \\
\Sgw= N_0 \mbox{diag}\left\lbrace (1+A^2|h_2[1]|^2),\cdots,(1+A^2|h_2[N]|^2) \right\rbrace
\end{gather}
as the covariance matrices of $\hb_1$ and $\w$, respectively.

Based on \eqref{eq:pdfY}, the maximum likelihood (ML) detection would be given as
\begin{equation}
\label{eq:ML}
\hat{\s}=\arg \max \limits_{\s \in \CpN} \left\lbrace \underset{\hb_2}{\E} \left\lbrace
\frac{1}{\pi^N \mathrm{det}\{\Sgy\}} \exp\left( -\y^H \Sgy^{-1} \y \right)
\right\rbrace \right\rbrace.
\end{equation}
where $\hat{\s}= \left[\; \hat{s}[1],\cdots, \hat{s}[N] \; \right]^t$.
As it can be seen, the ML metric needs the expectation over the distribution of $\hb_2$, which does not yield a closed-form expression. %Certain assumptions can be made to simplify the decision metric. For instance, if the RD link is very strong, the effect of the RD channel can be ignored to give a decision metric without any expectation. However, because of the dependence of $\Sgy$ to $\s$, one still needs an exhaustive search over all possible data sequences.
As an alternative, it is proposed to use the following modified decision metric:
\begin{equation}
\label{eq:ML-Modified}
\hat{\s}=\arg \max \limits_{\s \in \CpN} \left\lbrace
\frac{1}{\pi^N \mathrm{det}\{\overline{\Sg}_{\y}\}} \exp\left( -\y^H \overline{\Sg}_{\y}^{-1} \y \right)
\right\rbrace
\end{equation}
where
\begin{multline}
\label{eq:bSig_y}
\overline{\Sg}_{\y}=\underset{\hb_2}{\E} \{ \Sgy \}=A^2P_0 \dgs \Sg_{\hb} \dgsc\\+(1+A^2\sigma_2^2)N_0 \Ib_N
%A^2P_0 \dgs \Rb_{\hb} \dgsc+(1+A^2\sigma_2^2)N_0 \dgs \dgsc \\
=\dgs\; \C \; \dgsc
\end{multline}
with
\begin{equation}
\label{eq:C}
\C=A^2P_0  \Sg_{\hb} +(1+A^2\sigma_2^2)N_0\Ib_N
\end{equation}
\begin{multline}
\label{eq:R_h}
\Sg_{\hb}=\E\left\lbrace \dgh \Sg_{\hb_1} \dghc \right\rbrace=\\
\mathrm{toeplitz} \{ \varphi_1(0)\varphi_2(0),\cdots,\varphi_1(N-1)\varphi_2(N-1) \}.
\end{multline}
{Using Jensen's inequality \cite{jensen1}, iterative expectations and the concavity property of Eq. \eqref{eq:pdfY} over $\hb_2$, it can be shown that the function in the argument of the max function in Eq. \eqref{eq:ML-Modified} is an upper bound of that in Eq. \eqref{eq:ML}. In general, using the modified decision metric leads to a poorer detection error probability as compared to using the ML metric. Nevertheless, it will be shown by simulation results that nearly identical performance to that obtained with the optimal ML metric can be achieved with the modified metric.}

Using the rule $\det \{\A\B\}=\det\{\B\A\}$, the determinant in \eqref{eq:ML-Modified} is no longer dependent to $\s$ and the modified decision metric can be further simplified as
\begin{multline}
\label{eq:ML-simp}
\hat{\s}=\arg \min \limits_{\s\in \CpN} \left\lbrace \y^H \overline{\Sg}_{\y}^{-1} \y \right\rbrace\\
=\arg \min \limits_{\s\in \CpN} \{\y^H \dgs \C^{-1} \dgsc \y \}.
%\\
%=\arg \min \limits_{\s \in \CpN} \left\lbrace (\diag\{\y\} \s^*)^H \Lb\Lb^H \diag\{\y\} \s^* \right\rbrace\\
%=\arg \min \limits_{\s\in \CpN} \left\lbrace  \|\U \s \|^2 \right\rbrace.
\end{multline}
Next using the Cholesky decomposition of $\C^{-1}=\Lb \Lb^H$ gives
\begin{multline}
\label{eq:ML-simp2}
\hat{\s}=\arg \min \limits_{\s \in \CpN} \left\lbrace (\diag\{\y\} \s^*)^H \Lb\Lb^H \diag\{\y\} \s^* \right\rbrace\\
=\arg \min \limits_{\s\in \CpN} \left\lbrace  \|\U \s \|^2 \right\rbrace,
\end{multline}
where $\U=(\Lb^H \diag\{\y\})^*$.

The minimization in \eqref{eq:ML-simp2} can then be solved using sphere decoding described in \cite{MSDSD-L} to find $N-1$ information symbols. The MSD algorithm adapted for D-DH relaying is summarized in \emph{Algorithm I}. The complexity of the detection process is similar to that of point-to-point (P2P) communications, considered in \cite{MSDSD-L,msd-hassibi}. Although, here, the second-order statistics of two channels are required, in contrast to one channel in P2P communications. It should be mentioned that steps 1 to 3 are performed once, whereas steps 4 to 6 will be repeated for every $N$ consecutive received symbols. Also, the processed windows overlap by one symbol, i.e., the observation window of length $N$ moves forward by $N-1$ symbols at a time. Since, one symbol is taken as the reference, the output of the detection process contains $N-1$ detected symbols.

In the next section, the error analysis of the developed multiple-symbol detection is presented.

\begin{table}[thb!]
\renewcommand{\arraystretch}{2}
\label{tb:msdsd}
\centering
\begin{tabular}{l}
\hline
\bf Algorithm 1: MSD-DH \\
\hline
{\bf Input:} $\sigma_1^2,\sigma_2^2, f_1, f_2, A, P_0/N_0, M, N, \y$ \\
{\bf Output:} $\hat{v}[k],\quad k=1,\cdots,N-1$ \\
\hline
1: Find $\Rb_{\hb}$ from \eqref{eq:R_h} \\
2: Find $\C$ from \eqref{eq:C} \\
3: Find $\Lb$ from $\C^{-1}=\Lb\Lb^H$  \\
4: Find $\U=(\Lb^H \diag\{\y\})^*$ \\
5: Call function $\hat{\s}$=MSDSD ($\U$,$M$) \cite{MSDSD-L}\\
6: $\hat{v}[k]=\hat{s}^*[k] \hat{s}[k+1],\quad k=1,\cdots,N-1$\\
\hline
\end{tabular}
\end{table}

\subsection{Performance Analysis}
\label{subsec:msd-analysis}
Assume that vector $\s$ is transmitted and it is decoded as vector $\hat{\s}$. Based on the decision rule \eqref{eq:ML-simp}, an error occurs if
\begin{equation}
\label{eq:err1}
\y^H \diag\{\hat{\s}\} \C^{-1} \diag\{\hat{\s}^*\} \y \leq \y^H \diag\{{\s}\} \C^{-1} \diag\{{\s}^*\} \y,
\end{equation}
which can be simplified as
\begin{equation}
\label{eq:pep1}
\Delta=\y^H \Q \y \leq 0,
\end{equation}
with
\begin{equation}
\label{eq:Q}
\Q=\diag\{\hat{\s}-\s\} \C^{-1} \diag\{\hat{\s}^*-\s^*\}.
\end{equation}
Therefore, the PEP is defined as
\begin{equation}
\label{eq:msdPEP1}
P(\s \rightarrow \hat{\s})=P(\Delta\leq 0|\s,\hat{\s}).
\end{equation}
The above probability can be solved using the method of \cite{ber-biglieri} as
\begin{multline}
P(\Delta\leq 0|\s,\hat{\s})
\approx \\
 -\frac{1}{q} \sum \limits_{k=1}^{q/2} \left\lbrace c \Re[\Phi_{\Delta}(c+jc\tau_k)]+\tau_k \Im[\Phi_{\Delta}(c+jc\tau_k)] \right\rbrace
\end{multline}
where $\Phi_{\Delta}(\cdot)$ is the characteristic function of the random variable $\Delta$ and $\tau_k=\tan((2k-1)\pi/(2q))$. Also, the constant $c$ can be set equal to one half the smallest real part of the poles of $\Phi_{\Delta}(t)$ and $q=64$ gives enough accuracy for the approximation \cite{ber-biglieri}.

To proceed with computing \eqref{eq:msdPEP1}, the characteristic function of random variable $\Delta$ should be determined. Based on the modified decision rule \eqref{eq:ML-Modified}, conditioned on $\s$, $\y$ is circularly symmetric complex Gaussian vector with zero mean and covariance matrix $\bSigy$. Thus the characteristic function of the quadratic form $\Delta=\y^H \Q \y$ can be shown to be \cite{ch_quad}
\begin{equation}
\label{eq:phi_d}
\Phi_{\Delta}(t)=\frac{1}{\det\{\Ib_N+t\bSigy\Q\}}.
\end{equation}
By substituting \eqref{eq:phi_d} into \eqref{eq:msdPEP1}, the PEP of multiple-symbol detection is obtained. The BER can be obtained by the union bound of the PEP over the set of dominant errors as \cite{msdd-div}
\begin{equation}
\label{eq:msd-Pb}
P_{\bb}^{\MSD}\approx \frac{w}{\log_2(M)(N-1)} P(\s \rightarrow \hat{\s}),
\end{equation}
where $w$ is the sum of Hamming distances between the bit streams corresponding to the dominant errors. In the dominant errors, the detected vector $\hat{\s}$ is different than the transmitted vector $\s$ by only one closest symbol. For $M$-PSK, without loss of generality one can set
\begin{gather}
\s=[\;1,\cdots,1,1\;], \\
\hat{\s}=[\; 1,\cdots,1,\exp(j2\pi/M)\;]
\end{gather}
to find $P(\s\rightarrow \hat{\s})$. Also, with Gray-mapping \cite{msdd-div}
\begin{equation}
\label{eq:w}
w=\left\lbrace
\begin{matrix}
2(N-1), & \mbox{if} \; M=2 \\
4(N-1), & \mbox{if} \; M>2.
\end{matrix} \right.
\end{equation}

\section{Simulation Results}
\label{sec:sim}
In this section the dual-hop relay network under consideration is simulated for various channel qualities using both CDD and MSD to verify the analysis.

Information bits are differentially encoded with either BPSK ($M=2$) or QPSK ($M=4$) constellations. Note that, $\left\lbrace a=0,\; b=\sqrt{2}\right\rbrace$ and $\left\lbrace a=\sqrt{2-\sqrt{2}},\; b=\sqrt{2+\sqrt{2}}\right\rbrace$ are obtained for DBPSK and DQPSK, respectively \cite{unified-app}. The amplification factor at Relay is fixed to $A=\sqrt{P_1/(P_0\sigma_1^2+N_0)}$ to normalize the average relay power to $P_1$. Also, $N_0=1$ is assumed.

Based on the fading powers of the channels, three scenarios would be considered as: symmetric channels with $\sigma_1^2=1,\sigma_2^2=1$, non-symmetric channels with strong SR channel $\sigma_1^2=10,\sigma_2^2=1,$ and non-symmetric channels with strong RD channel $\sigma_1^2=1,\sigma_2^2=10$. The fading powers are listed in Table~\ref{table:sig12}.

\begin{figure}[htb!]
\psfrag {PAF} [t][] [1]{$\varrho=\frac{P_0}{P}$}
\psfrag {BER} [] [] [1] {BER}
\psfrag {sig12=110} [] [] [1] {$\quad \qquad \sigma_1^2=1,\sigma_2^2=10$}
\psfrag {sig12=101} [] [] [1] {$\quad \qquad\sigma_1^2=10,\sigma_2^2=1$}
\psfrag {sig12=11} [] [] [1] {$\quad \qquad\sigma_1^2=1,\sigma_2^2=1$}
\psfrag {P/N0} [] [] [1] {$P/N_0=[30,35,40]$ dB}
\centerline{\epsfig{figure={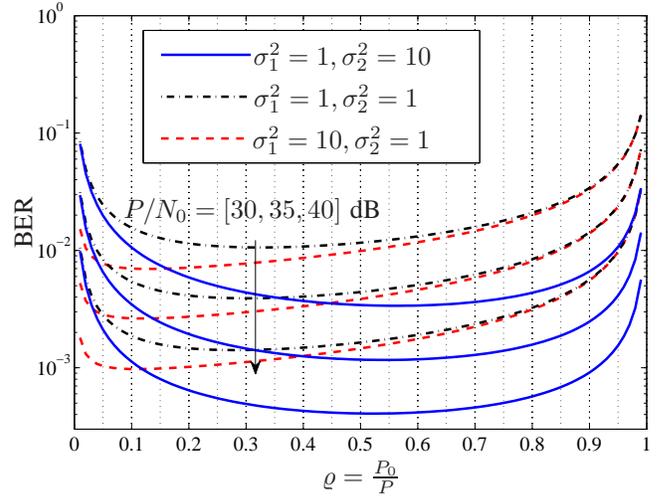},width=8.5cm}}
\caption{BER as a function of power allocation factor for various channel variances and $P/N_0=[ 30,35,40 ]$ dB from top to bottom, using DQPSK.}
\label{fig:pw_m4_all}
\end{figure}

\begin{table}[!ht]
\begin{center}
\caption{Fading powers and corresponding optimum power allocation factors.}
\label{table:sig12}
\begin{tabular}{cc|c|c|}
\cline{2-3}
&\multicolumn{1}{|c|}  {$[\sigma_1^2, \sigma_2^2]$} & $\varrho_{\opt}$   \\ \cline{1-3}
\multicolumn{1}{ |c| } {Symmetric} & {$[1,1]$} & 0.30   \\
\cline{1-3}
\multicolumn{1}{ |c| }{Strong SR} & {$[10,1]$} & 0.12   \\
\cline{1-3}
\multicolumn{1}{ |c| }{Strong RD} & {$[1,10]$} & 0.54  \\
\cline{1-3}
\end{tabular}
\end{center}
\end{table}
First, the BER expression of CDD is used to optimize the power allocation between Source and Relay in the network. The optimization problem aims to minimize the BER for a given total power $P=P_0+P_1$. Let $P_0=\varrho P$, where $\varrho$ is the power allocation factor. Then $P_1=(1-\varrho)P$ and from \eqref{eq:A}, $A=\sqrt{P(1-\varrho)/(\varrho P \sigma_1^2+N_0)}$. By substituting $P_0$ and $A$ into \eqref{eq:Pb}, the optimization problem is to find the value of $\varrho$ for which $P_{\mathrm{b}}(E)$ is minimized. Since the best performance is obtained in slow-fading channels, $\alpha=1$ would be set in \eqref{eq:Pb}. Then, the minimization would be performed by a numerical exhaustive search.
%For the analytical approach, it is easy to see that the minimization does not depend on $\theta$. Thus minimizing $P_{\mathrm{b}}(E)$ is equivalent to minimizing $J(\theta)$, for any value of $\theta,$ say $\theta=\pi/2$. Therefore, the optimized power allocation is obtained by minimizing
%\begin{equation}
%\label{eq:Jtilde}
%\tilde{J}=\tilde{b}_3+\frac{\tilde{b}_3}{\sigma_2^2} (\tilde{b}_1-\tilde{b}_2) e^{\frac{\tilde{b}_2}{\sigma_2^2}} E_1\left( \frac{\tilde{b}_2}{\sigma_2^2}\right)
%\end{equation}
%where $\tilde{b}_3=\tilde{b}_2/\tilde{b}_1$, $\tilde{b}_1=1/A^2$ and $\tilde{b}_2=2/(2A^2+q(\pi/2)A^2\rho_1)$. It can be seen that for large $P$ or $P_0$, $\tb_1-\tb_2\approx \tb_1 $ and then $\tb_3(\tb_1-\tb_2)\approx \tb_2$. On the other hand, for small $x=\tb_2/\sigma_2^2$ one has the following approximation \cite[eq.5.1.20]{integral-tables}
%$$
%\label{eq:exE1x}
%e^x E_1(x)\approx \log\left( \frac{1}{x}\right).
%$$
%Thus, $\tilde{J}$ can be approximated as
%\begin{equation}
%\label{eq:Jtilde_approx}
%\tilde{J} \approx \tb_3 + \frac{\tb_2}{\sigma_2^2} \log\left( \frac{\sigma_2^2}{\tb_2}\right)
%\end{equation}
%Then, finding the minimum of \eqref{eq:Jtilde_approx} over the single variable $\varrho$. can be easily done by solving the derivative of \eqref{eq:Jtilde_approx} respect to $\varrho.$
The obtained optimum power allocations are listed in Table~\ref{table:sig12} for $P/N_0=35$ dB. Note that $P$ is the total power in the network and is divided between Source and Relay. It will be seen that, to achieve a low BER around $10^{-3}$ or $10^{-4}$, such a high value would be needed. In addition, the BER curves obtained from the exhaustive search are plotted versus $\varrho$ in Figure~\ref{fig:pw_m4_all} for $P/N_0=30,\;35,\;40$ dB and when DQPSK is employed. Similar results would be obtained when DBPSK is employed.  The results in Table~\ref{table:sig12} and Figure~\ref{fig:pw_m4_all} show that for symmetric and strong SR channels, more power should be allocated to Relay than Source and the BER is minimized at $\varrho_{\mathrm{opt}}\approx 0.3$ and $\varrho_{\mathrm{opt}}\approx 0.1$, for symmetric and strong SR channels, respectively. When the RD channel becomes stronger than the SR channel, the system would benefit more from an equal power allocation and the BER is minimized at $\varrho_{\mathrm{opt}}\approx 0.5$.

In all simulations, the channel coefficients, $h_1[k]$ and $h_2[k]$, are generated based on the simulation method of \cite{ch-sim}. This simulation method was developed to generate channel coefficients that are correlated in time. The amount of time-correlation is determined by the normalized Doppler frequency of the underlying channel, which is a function of the speed of the vehicle, carrier frequency and symbol duration. For fixed carrier frequency and symbol duration, a higher vehicle speed leads to a larger Doppler frequency and less time-correlation.

\begin{figure}[htb!]
\psfrag {f1 changes} [t][] [.8]{$f_1$ changes}
\psfrag {f12 change} [b][] [.8]{$f_1\&f_2$ change}
\psfrag {f} [t][] [1]{fade rate}
\psfrag {BER} [] [] [1] {Error Floor}
\psfrag {Analysis} [] [r] [.8] {Analysis \eqref{eq:PEP-floor}}
\psfrag {correlation} [] [] [1] {Autocorrelation}
\centerline{\epsfig{figure={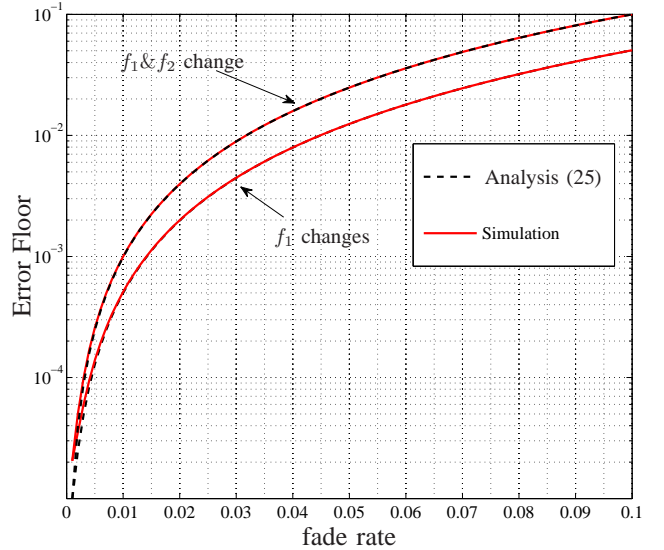},width=8.5cm}}
\caption{Theoretical and simulation values of error floor vs. fading rate, $M=2, \varrho=0.3$, $[\sigma_1^2,\sigma_2^2]=[1,1]$. }
\label{fig:ef_fd}
\end{figure}
To get a better understanding about the fading rate values and also verifying the analysis, the error floor values of CDD are obtained from both expression \eqref{eq:PEP-floor} and simulation at a high transmit power for a wide range of fading rates and plotted in Figure~\ref{fig:ef_fd}.
%Other parameters are $M=2$, with symmetric channels and $\varrho=0.25, P/N_0=60$dB.
In the figure, the lower graph corresponds to the case that only $f_1$ changes and $f_2=0.001$. In the upper graph both $f_1$ and $f_2$ change. Clearly, the error floors from both analysis and simulation tight together. For small values of fading rate around 0.001, the error floor is very low, around $10^{-5}$, but increases toward $10^{-3 }$ with increasing the fading rate to $0.01$. The fading rate around $0.01$ can be regarded as the margin beyond which the channels become fast-fading. As also seen, the error floor values are higher when both Doppler values change.

Based on the previous observation and the normalized Doppler frequencies of the two channels, different cases can be considered. In Case I, it is assumed that all nodes are fixed or slowly moving so that both channels are slow-fading with the normalized Doppler values of $f_1=.001$ and $f_2=.001$. In Case II, it is assumed that Source is moving so that the SR channel is fast-fading with $f_1=.01$. On the other hand, Relay and Destination are fixed and the RD channel is slow-fading with $f_2=.001$. In Case III, it is assumed that both Source and Relay are moving so that both the SR and RD channels are fast-fading with $f_1=.02$ and $f_2=.01$, respectively.
The normalized Doppler values are shown in Table \ref{table:f1f2}. %Also, a snapshot of realizations of the cascaded channel and their auto-correlation values in the three cases are plotted in Figure~\ref{fig:doub-ch}.
\begin{table}[!ht]
\begin{center}
\caption{Three fading scenarios.}
\label{table:f1f2}
  \begin{tabular}{ |c | c| c|c|  }
    \hline
				& $f_1$ & $f_2$ & Channels status   \\ \hline\hline
{Case I }& 0.001 & 0.001  & both are slow-fading \\ \hline
{Case II } & 0.01 & 0.001& SR is fast-fading  \\ \hline
{Case III } & 0.02 & 0.01 & both are fast-fading  \\
    \hline
  \end{tabular}
\end{center}
\end{table}
%\vspace*{-0.5cm}

%\begin{figure}[htb!]
%\psfrag {time} [][] [1]{$k$}
%\psfrag {h} [] [] [1] {$|h[k]|$}
%\psfrag {correlation} [] [] [1] {Autocorrelation}
%\centerline{\epsfig{figure={figure/dch_scs.eps},width=8.5cm}}
%\caption{Snapshot of realizations of the cascaded channel and corresponding autocorrelations in different cases.}
%\label{fig:doub-ch}
%\end{figure}

\begin{figure}[htb!]
\psfrag {P(dB)} [t][b] [1]{$(P/N_0)$ (dB)}
\psfrag {BER} [] [] [1] {BER}
\centerline{\epsfig{figure={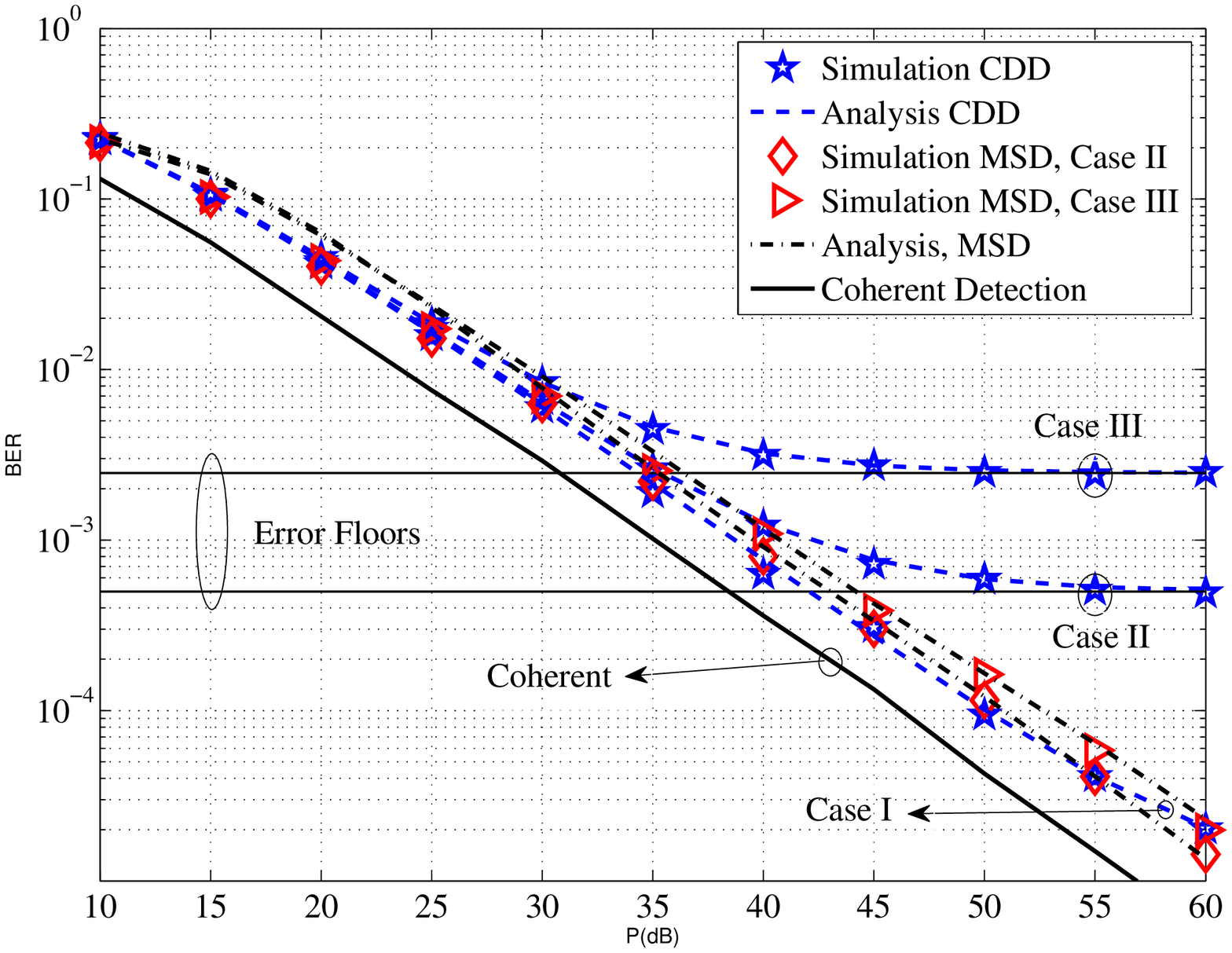},width=8.5cm}}
\caption{Theoretical and simulation BER of a D-DH relaying in different fading cases and $[\sigma_1^2,\sigma_2^2]=[1,1]$ using DBPSK and CDD ($N=2$) and MSD ($N=10$).}
\label{fig:ber_m2_sg11}
\end{figure}

\begin{figure}[htb!]
\psfrag {P(dB)} [t][b] [1]{$(P/N_0)$ (dB)}
\psfrag {BER} [] [] [1] {BER}
\centerline{\epsfig{figure={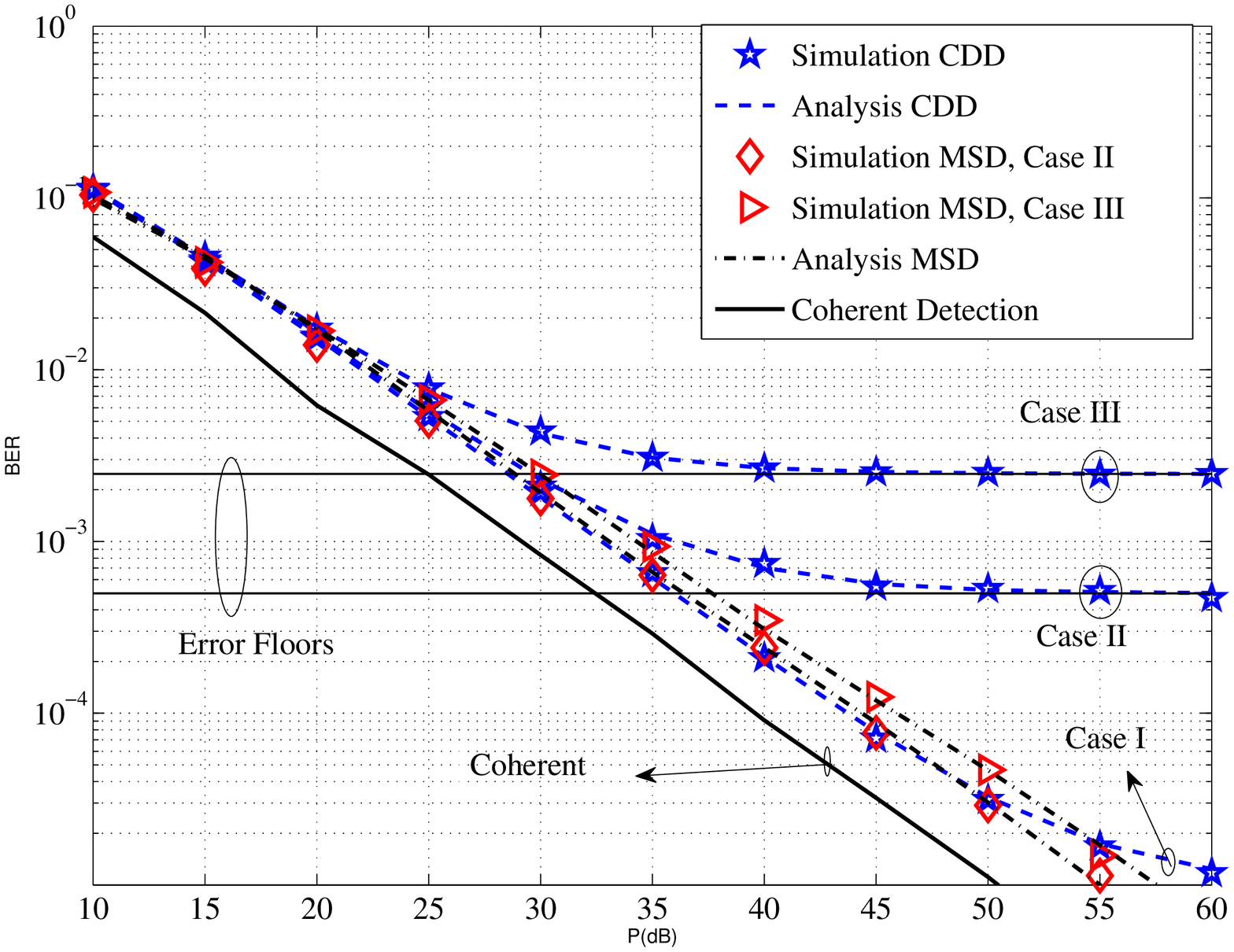},width=8.5cm}}
\caption{Theoretical and simulation BER of a D-DH relaying in different fading cases and $[\sigma_1^2,\sigma_2^2]=[1,10]$ using DBPSK and CDD ($N=2$) and MSD ($N=10$).}
\label{fig:ber_m2_sg110}
\end{figure}

\begin{figure}[htb!]
\psfrag {P(dB)} [t][b] [1]{$(P/N_0)$ (dB)}
\psfrag {BER} [] [] [1] {BER}
\centerline{\epsfig{figure={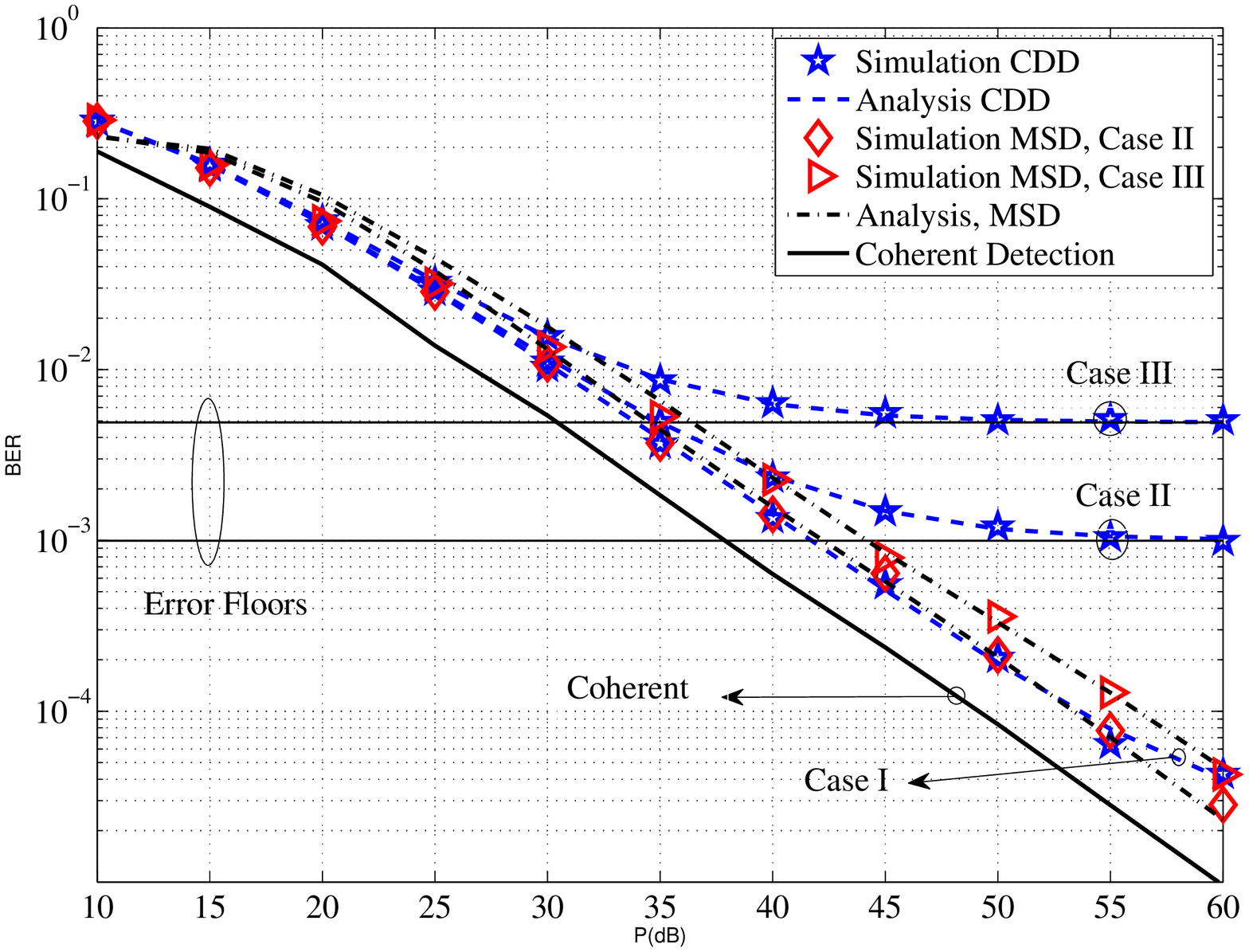},width=8.5cm}}
\caption{Theoretical and simulation BER of a D-DH network in different fading cases and $[\sigma_1^2,\sigma_2^2]=[1,1]$ using DQPSK and CDD ($N=2$) and MSD ($N=10$).}
\label{fig:ber_m4_sg11}
\end{figure}

\begin{figure}[htb!]
\psfrag {P(dB)} [t][b] [1]{$(P/N_0)$ (dB)}
\psfrag {BER} [] [] [1] {BER}
\centerline{\epsfig{figure={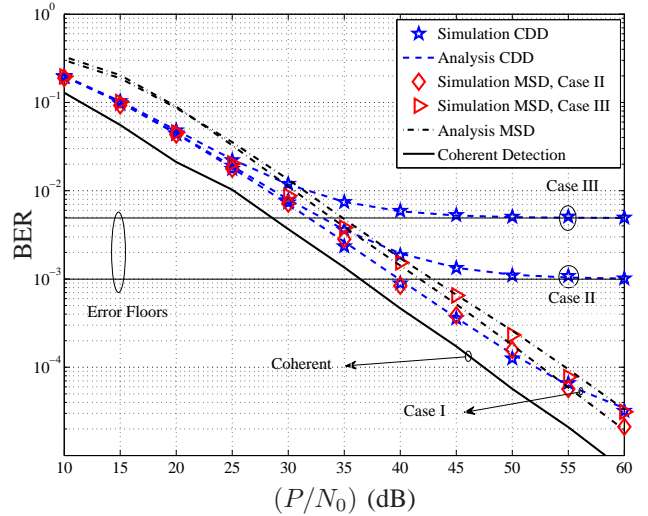},width=8.5cm}}
\caption{Theoretical and simulation BER of a D-DH relaying in different fading cases and $[\sigma_1^2,\sigma_2^2]=[10,1]$ using DQPSK and CDD ($N=2$) and MSD ($N=10$).}
\label{fig:ber_m4_sg101}
\end{figure}

Based on the obtained optimum power allocation values in Table~\ref{table:sig12}, the simulated BER values using the CDD are computed for all cases and different channel variances and are plotted versus $P/N_0$ in Figs.~\ref{fig:ber_m2_sg11}-\ref{fig:ber_m4_sg101}, for DBPSK and DQPSK modulations. Also, the simulated BER values using coherent detection are computed for Case I (slow-fading) and plotted in Figs.~\ref{fig:ber_m2_sg11}-\ref{fig:ber_m4_sg101}. Figures~\ref{fig:ber_m2_sg11} and \ref{fig:ber_m2_sg110} correspond to symmetric channels and strong RD channel, respectively, while Figures~\ref{fig:ber_m4_sg11} and \ref{fig:ber_m4_sg101} correspond to symmetric channels and strong SR channel, respectively. %In addition, the simulated outage probability for $\gth=0$ and 5 dB using the CDD are also computed for all cases and different channel variances. They are plotted versus $P/N_0$ in Figs.~\ref{fig:pout_m2_scs_sg101_gths}-\ref{fig:pout_m4_scs_sg110_gths}, for DBPSK with strong SR channel and DQPSK with strong RD channel, respectively.
On the other hand, the corresponding theoretical BER values (for all cases) and the error floors (cases II\&III) for the CDD are computed from (\ref{eq:Pb}) and \eqref{eq:PEP-floor} (with the corresponding variances) and plotted in Figs.~\ref{fig:ber_m2_sg11}-\ref{fig:ber_m4_sg101}.

As can be seen in Fig.~\ref{fig:ber_m2_sg11}, in Case I with symmetric channels, the BER is monotonically decreasing with $(P/N_0)$ and it is consistent with the theoretical values in (\ref{eq:Pb}). Approximately, 3 dB performance loss is seen between coherent and non-coherent detection in this case. The error floor in this case does not practically exist. In Case II, which involves one fast-fading channel, this phenomenon starts earlier, around $35$ dB, and leads to an error floor at $5\times 10^{-4}$, which can also be predicted from (\ref{eq:PEP-floor}). The performance degradation is much more severe after $25$ dB in Case III since both channels are fast-fading, which leads to an error floor at $3\times 10^{-3}$. Similar behaviours can be seen in Figs.~\ref{fig:ber_m2_sg110}-\ref{fig:ber_m4_sg101} under other channel variances or DQPSK modulation. As is clearly seen from Figs.~\ref{fig:ber_m2_sg11}-\ref{fig:ber_m4_sg101}, the simulation results verify our theoretical evaluations. Specifically, the error floors in Case II and Case III do not depend on the channel variances.

Given the poor performance of the two-symbol detection in Cases II and III, MSD-DH algorithm is applied which takes a window of $N=10$ symbols for detection. The BER results of the MSD-DH algorithm are also plotted in Figures~\ref{fig:ber_m2_sg11} -\ref{fig:ber_m4_sg101} with points (different legends). Also, theoretical BER of MSD in Cases II and III are obtained from \eqref{eq:msd-Pb} and plotted in the figures with dash-dot lines. Since the best performance is achieved in the slow-fading environment, the performance plot of Case I can be used as a benchmark to see the effectiveness of MSD-DH. It can be seen that the MSD-DH is able to bring the performance of the system in Case II and Case III very close to that of Case I. Moreover, the analytical results of MSD are consistent with the simulation results and they match well in high SNR region.

%\begin{figure}[htb!]
%\psfrag {P(dB)} [][b] [1]{$(P/N_0)$ (dB)}
%\psfrag {Pout} [] [] [1] {$P_{\mathrm{out}}$}
%\centerline{\epsfig{figure={figure/pout_m2_scs_sg101_gths.eps},width=8.5cm}}
%\caption{Outage probability of a D-DH network in different fading cases and $\sigma_1^2=10,\sigma_2^2=1$ using DBPSK and $\gth=$ 0 and 5 dB .}
%\label{fig:pout_m2_scs_sg101_gths}
%\end{figure}
%
%\begin{figure}[htb!]
%\psfrag {P(dB)} [][b] [1]{$(P/N_0)$ (dB)}
%\psfrag {Pout} [] [] [1] {$P_{\mathrm{out}}$}
%\centerline{\epsfig{figure={figure/pout_m4_scs_sg110_gths.eps},width=8.5cm}}
%\caption{Outage probability of a D-DH network in different fading cases and $\sigma_1^2=1,\sigma_2^2=10$ using DQPSK and $\gth=$ 0 and 5 dB .}
%\label{fig:pout_m4_scs_sg110_gths}
%\end{figure}

%\vspace*{-0.35cm}

\section{Conclusion}
\label{sec:con}
Differential AF relaying for a dual-hop transmission has been studied for time-varying Rayleigh fading channels. {Performance analysis was carried out to show that, though being simple, the conventional two-symbol differential detection (CDD) suffers from an error floor in fast-fading channels. Such analysis is useful to predict the error floor phenomenon and control its value by suitably changing system parameters such as the symbol duration. Furthermore, a multiple-symbol detection (MSD) scheme was proposed and its performance is also theoretically analyzed. The theoretical and simulation results demonstrated that, at the expense of higher detection complexity, the proposed MSD scheme can significantly improve the poor performance of the CDD in fast-fading channels. Since channel estimation required for coherent detection becomes severely inaccurate in fast-fading environment, the proposed MSD scheme offers an attractive option to trade complexity for system performance of dual-hop transmission in fast-fading channels.}

%\balance
\bibliographystyle{IEEEbib}
%\bibliography{h:/latex/references}
\bibliography{references}

\end{document}